\documentclass[11pt]{article}
\usepackage[dvips]{color}
\usepackage{amssymb}
\usepackage{amsmath}
\usepackage{amscd}
\usepackage{latexsym}
\usepackage{ifthen}
\catcode`@=11
\renewcommand{\section}{\@startsection{section}{1}{0pt}{\medskipamount}
{\medskipamount}{\large\bf}}
\catcode`@=12

\def\b{\beta}
\def\g{\gamma}

\def\ve{\varepsilon}

\def\h{\eta}

\def\l{\lambda}
\def\m{\mu}

\def\s{\sigma}

\def\P{\Phi}
\def\L{\Lambda}

\newcommand{\C}{\mathbb C}

\newcommand{\Z}{\mathbb Z}

\newcommand{\Ical}{{\cal I}}
\newcommand{\Pcal}{{\cal P}}
\newcommand{\Qcal}{{\cal Q}}
\newcommand{\Xcal}{{\cal X}}
\newcommand{\Ucal}{{\cal U}}
\newcommand{\Acal}{{\cal A}}
\newcommand{\Tcal}{{\cal T}}
\newcommand{\Wcal}{{\cal W}}

\def\>{\rangle}
\def\<{\langle}
\def\={\ =\ }
\def\+{\dagger}
\def\e{\textrm{e}}
\def\i{\textrm{i}}
\def\Ne1{$N\,{=}\,1$}
\def\N2{$N\,{=}\,2$}
\def\Ng4{$N\,{=}\,4$}
\def\pa{\partial}

\def\sfrac#1#2{{\textstyle\frac{#1}{#2}}}
\def\Kpmb#1{\setbox0=\hbox{${#1}$}   \kern-.025em\copy0\kern-\wd0
    \kern.05em\copy0\kern-\wd0     \kern-.025em\raise.0433em\box0 }

\newcommand{\ov}[1]{\overline{#1}}
\newcommand{\wt}[1]{\widetilde{#1}}
\newcommand{\wh}[1]{\widehat{#1}}
\newcommand{\lb}{\bar{\l}}
\newcommand{\mb}{\bar{\mu}}

\newcommand{\Tb}{\ov{T}}
\newcommand{\Tcb}{\ov{\Tcal}}

\newcommand{\At}{\wt{A}}
\newcommand{\Gt}{\wt{G}}

\newcommand{\Pst}{\wt{\Psi}}

\newcommand{\hh}{\wh{\h}_0}

\newcommand{\Phh}{\wh{\Phi}}
\newcommand{\Psh}{\wh{\Psi}}
\newcommand{\Ach}{\wh{\Acal}}
\newcommand{\Ahg}{\wh{\Acal}_g}
\newcommand{\Pch}{\wh{\Pcal}}
\newcommand{\Phg}{\wh{\Pcal}_g}
\newcommand{\Qch}{\wh{\Qcal}}
\newcommand{\Xch}{\wh{\Xcal}}

\newcommand{\Gp}{\ifthenelse{\boolean{mmode}}{{G^+}}{\mbox{$G^+\:$}}}
\newcommand{\Gtp}{\ifthenelse{\boolean{mmode}}{\mbox{$\Gt^+$}}{\mbox{$\Gt^+\:$}}}
\newcommand{\Gm}{\ifthenelse{\boolean{mmode}}{{G^-}}{\mbox{$G^-\:$}}}
\newcommand{\Gtm}{\ifthenelse{\boolean{mmode}}{\mbox{$\Gt^-$}}{\mbox{$\Gt^-\:$}}}

\begin{document}
\begin{titlepage}
\setcounter{page}{0}
\begin{flushright}
hep-th/0209186\\
ITP--UH--22/02\\
\end{flushright}

\vskip 2.0cm

\begin{center}

{\Large\bf On Nonperturbative Solutions of Superstring Field Theory}

\vspace{14mm}

{\large Alexander Kling, Olaf Lechtenfeld, Alexander D.\ Popov$\,^*$
	\ and \ Sebastian Uhlmann}
\\[5mm]
{\em Institut f\"ur Theoretische Physik  \\
Universit\"at Hannover \\
Appelstra\ss{}e 2, 30167 Hannover, Germany }\\[2mm]
{Email: kling, lechtenf, popov, uhlmann@itp.uni-hannover.de}

\end{center}

\vspace{2cm}

\begin{abstract}
\noindent
Nonperturbative solutions to the nonlinear field equations in the NS~sector 
of cubic as well as nonpolynomial superstring field theory can be obtained 
from a linear equation which includes a ``spectral'' parameter~$\l$ and a 
coboundary operator~$Q(\l)$.  We borrow a simple ansatz from the dressing 
method (for generating solitons in integrable field theories) and show that 
classical superstring fields can be constructed from any string field~$T$ 
subject merely to $Q(\l)T=0$.  Following the decay of the non-BPS D9 brane 
in IIA theory and shifting the background to the tachyon vacuum, we repeat 
the arguments in vacuum superstring field theory and outline how to compute 
classical solutions explicitly. 
\end{abstract}

\vfill

\textwidth 6.5truein
\hrule width 5.cm
\vskip.1in

{\small
\noindent ${}^*$
On leave from Bogoliubov Laboratory of Theoretical Physics, JINR,
Dubna, Russia}

\end{titlepage}

%%%%%%%%%%%%%%%%%%%%%%%%%%%%%%%%%%%%%%%%%%%%%%%%%%%%%%%%
\noindent
{\bf 1.~Introduction.\ }
Sen's conjectures (for a review see, e.~g., \cite{Sen:1999mg,
Ohmori:2001am} and references therein) on tachyon condensation
have sparked considerable activity in open string field theory.
Even though there is no tachyon in the superstring spectrum
(on flat spacetime) the decay of unstable non-BPS D9-branes
in type IIA theory is due to the GSO$(-)$ NS tachyon excitation.
The dynamics of this transition from the D9-brane vacuum
to the tachyon vacuum involves nonperturbative classical superstring
configurations. Thus, the task is to solve the NS string field
equations of motion, either in Witten's cubic~\cite{Witten:1986qs}
or in Berkovits' nonpolynomial~\cite{Berkovits:1995ab} formulation.
Although some progress has been made in this direction~%
\cite{Marino:2001ny,Arefeva:2002mb,Ohmori:2002ah,Ohmori:2002kj} 
we still do not have an exact solution at hand as of today.

Some time ago it was shown by two of us~\cite{Lechtenfeld:2000qj}
that Berkovits' string field theory is integrable in the sense that
its equation of motion derives from a system of linear equations.
Clearly, one should take advantage of this fact and try to bring to
application the powerful solution-generating technology available
for integrable equations. The goal of this letter is to initiate
such a program, based on the ideas presented
in~\cite{Lechtenfeld:2002cu}.

In analogy with gauge field theory, we write down a linear system for
cubic as well as nonpolynomial open superstring field theory
(in the NS sector) by introducing an auxiliary string field~$\Psi(\l)$ 
depending on a ``spectral'' parameter $\l\in\C P^1$.
A single-pole ansatz for $\Psi(\l)$ leads to a hermitian projector, 
whose building block is merely subject to a linear equation 
which can be solved in generality. From it
all string fields can be reconstructed. Employing dressing
transformations analogous to those in noncommutative field theories~%
\cite{Lechtenfeld:2001aw}, we shift the background to the tachyon vacuum and
propose a linear equation which governs classical vacuum superstring field 
theory. As a simple example, the supersliver~\cite{Marino:2001ny,Arefeva:2001ke}
is based on a trivial solution to this equation. Finally, we propose
a strategy to reconstruct classical superstring fields from
their building blocks in more detail by taking advantage of the Moyal 
formulation for superstring field theory.

\noindent
{\bf 2.~Zero-curvature and linear equations for string fields.\ }
In cubic open bosonic string field theory~\cite{Witten:1985cc}, the
equation of motion for the string field~$A$ has a zero-curvature form,
\begin{equation} \label{F0b}
  F(A)\=Q A+A^2\=(Q+A)^2\=0 \quad,
\end{equation}
where $Q$ denotes the BRST operator (a nilpotent derivation) and
Witten's star product is implicit in all string field products.
For any string field~$A$ one may look for solutions of the linear equation
\begin{equation} \label{lin}
  (Q+A)\,\Psi\=0
\end{equation}
on an auxiliary string field~$\Psi$ possibly carrying some internal
indices. Equation~(\ref{F0b}) is the compatibility condition of the
linear equation~(\ref{lin}). If we let~$\Psi$ take values in the
Chan-Paton group, then from~(\ref{lin}) one may obtain solutions 
of~(\ref{F0b}) via $\ A=\Psi Q\Psi^{-1}\ $ which are, however, 
pure gauge configurations.  
The cohomology of~$Q$ captures all other solutions.

This situation may change when a parametric dependence is introduced:
Let $(Q,A,\Psi)\to(Q(\l),A(\l),\Psi(\l))$ with $\l\in\C P^1$.
We demand $Q(\l)$ and $A(\l)$ to be linear in~$\l$,\footnote{Formally
$A(\l)$ is a section of the bundle ${\cal O}(1)$ over $\C P^1$ with
values in the string field Hilbert space ${\cal H}$, and $Q(\l)$ can
be considered as an End$({\cal H})$-valued section of this bundle.}
\begin{equation}
  A(\l)\=a+\l A \qquad\textrm{and}\qquad
  Q(\l)\=\h_0+\l Q \qquad\textrm{with}\qquad
  \h_0^2=Q^2=\h_0\,Q+Q\,\h_0=0 \quad.
\end{equation}
In other words, we extend the string configuration space,
thereby adding a second string field~$a$
and a second BRST-like operator~$\h_0$.
This case arises for a one-parameter family of \N2\ superconformal algebras
embedded into a small \Ng4\ algebra and their string field realizations~%
\cite{Berkovits:1994vy, Berkovits:1999im}. The extended zero-curvature
condition
\begin{equation} \label{F0l}
  F\bigl(A(\l)\bigr)\=\bigl( Q(\l)+A(\l) \bigr)^2 \=
  \bigl( \h_0 a + a^2 \bigr) + \l\bigl( \h_0 A + Q a
  + \{ A,a \} \bigr) + \l^2 \bigl( QA+A^2 \bigr) \=0
\end{equation}
is the compatibility condition of the associated linear equation
\begin{equation}
  \bigl( Q(\l) + A(\l) \bigr)\,\Psi(\l) \=0 \quad .
\end{equation}
If $\Psi(\l)$ is group-valued, it follows that
$\ a+\l A=\Psi(\l)\bigl(\h_0+\l Q\bigr)\Psi(\l)^{-1}$.
As was shown in~\cite{Lechtenfeld:2000qj,Lechtenfeld:2002cu}, 
this equation yields nontrivial solutions to the equations of motion for
$a$ and~$A$.

Exploiting the gauge freedom in~(\ref{F0l}) allows one to
gauge away~$a$. Then, the ensuing equations,
\begin{equation} \label{F0s}
  \h_0\,A\=0 \qquad\textrm{and}\qquad Q A+A^2\=0
\end{equation}
are the (NS-sector) equations of motion in Witten's cubic open
superstring field theory in the zero picture \cite{Preitschopf:fc,
Arefeva:1989cp}: Bosonizing the fermionic reparametrization ghosts,
$\g=\h\e^{\phi}$ and $\b=\e^{-\phi}\pa\xi$,
we take $\h_0$ above to be the zero mode of~$\h$,
which indeed is nilpotent and anticommutes with~$Q$.
Then, the first equation in~(\ref{F0s}) simply denies any $\xi_0$
content in~$A$ (originally defined in the large Hilbert space),
and the second one is the field equation in the small Hilbert space.\footnote{
Note that $Q$ and $\h_0$ act via (anti)commutator on world-sheet fields, or,
equivalently, via contour integration of the respective currents.}
Of course, all fields are now NS-sector open superstring fields.

The system~(\ref{F0s}) may be reduced further.\footnote{
For gauge theory the following goes back to Leznov and to Yang, respectively.}
Since both $\h_0$ and $Q$ have trivial cohomology in the large 
Hilbert space~${\cal H}$, we may either solve the first equation
or alternatively the second one:
\begin{equation} \label{leznov}
  A\=\h_0\,\Upsilon \qquad\Longrightarrow\qquad
  Q\,\h_0\,\Upsilon + (\h_0\,\Upsilon)^2 \=0 \quad,
\end{equation}
\begin{equation} \label{yang}
  A\=\e^{-\P}Q\e^{\P} \qquad\Longrightarrow\qquad
  \h_0 \bigl( \e^{-\P}Q\e^{\P} \bigr) \=0 \quad.
\end{equation}
Despite appearance, $A$ is {\it not\/} pure gauge (in the small Hilbert
space) unless $\ \h_0\e^\P=0$ \cite{Ohmori:2002ah}.
The second equation in~(\ref{yang}) is precisely Berkovits' nonpolynomial
equation of motion for the NS string field~$\P$.

All nonlinear superstring field equations,
i.~e. (\ref{F0s}), (\ref{leznov}) and (\ref{yang}),
follow from the zero-curvature equation~(\ref{F0l}) (with $a{=}0$).
Because both $Q$ and $\h_0$ have empty cohomology in the large Hilbert space
we can in fact construct all solutions from the associated linear system
\begin{equation} \label{linl}
  \bigl( Q + \sfrac{1}{\l}\h_0 + A \bigr)\,\Psi(\l) \=0
\end{equation}
for the string fields $A$ and~$\Psi(\l)$.\footnote{Formally $\Psi(\l)$
can be seen as an element of the space ${\cal H}\otimes\C[\l,\l^{-1}]$
carrying Chan-Paton labels.}
This equation is the key to generating classical superstring
configurations.

Of course, one always has the ``trivial'' $\l$-independent solution
\begin{equation}
  \Psi \= \e^{-\L} \quad\textrm{with}\quad \pa_\l \L=0
  \qquad\Longrightarrow\qquad
  \h_0\,\e^{-\L}\=0\=(Q+A)\,\e^{-\L}
\end{equation}
which leads to a pure gauge configuration $A_0=\e^{-\L}Q\e^{\L}$.
Since $\C P^1$ is compact, the $\l$ dependence of a nontrivial $\Psi(\l)$
cannot be holomorphic.
Hence, we consider a meromorphic~$\Psi(\l)$.
If we require its regularity for $\l\to0$ and for $\l\to\infty$,
then one may choose such a gauge that the asymptotics will relate $\Psi$
with the prepotentials $\P$ and~$\Upsilon$ as follows:\footnote{
$\Ical$ denotes the identity string field.}
\begin{equation}
  \Psi(\l)\ \longrightarrow\
  \begin{cases}
    \Ical - \l\,\Upsilon + O(\l^2)      & \textrm{for} \quad \l\to 0 \\
    \e^{-\P} + O(\sfrac{1}{\l}) & \textrm{for} \quad \l\to\infty
  \end{cases}
  \quad.
\end{equation}
Clearly, $e^{-\P}$, $\Upsilon$,
and $\ A=\Psi(\infty)Q\Psi(\infty)^{-1}=-\h_0\,\pa_\l\Psi(0)\ $
are computable once an appropriate $\Psi(\l)$ has been found.

\noindent
{\bf 3.~Single-pole ansatz and solutions.\ }
Let us employ the linear system~(\ref{linl}) to solve Witten's or
Berkovits' superstring field equations (in the NS sector).
In contrast to the non-parametric linear equation~(\ref{lin}),
the $\l$ dependence of~(\ref{linl}) imposes two constraints on~$\Psi(\l)$.
Firstly, isolating $A$ in~(\ref{linl}),
\begin{equation} \label{nopole1}
  A\= \Psi(\l) \bigl( Q + \sfrac{1}{\l}\h_0 \bigr) \Psi(\l)^{-1} \quad,
\end{equation}
we notice that the right-hand side must not depend on~$\l$, hence all its
poles must have vanishing residues.
Although the above expression is pure gauge from the point of view of
the $\l$-extended string configuration space,
the string field~$A$ is nontrivial on the small Hilbert space.
A second condition follows from the reality of the string fields.
To formulate it one must extend hermitian conjugation to an antilinear
mapping (which we denote by a bar) on the $\C P^1$ family of \N2\
superconformal algebras where it sends $\ Q\mapsto-\h_0\ $ and
$\ \h_0\mapsto Q\ $ but $\ \l\mapsto\lb\ $~\cite{Lechtenfeld:2002cu}. It can
be shown that the reality condition requires
\begin{equation} \label{nopole2}
  \e^{-\P}\= \Psi(\l)\,\ov{\Psi(-1/\lb)} \quad.
\end{equation}
Again, the poles on the right-hand side must be removable.

The simplest nontrivial solution displays a single pole in~$\l$,\footnote{
For more general multi-pole ans\"atze see \cite{Lechtenfeld:2002cu}.}
\begin{equation} \label{ansatz1}
  \Psi(\l)\=\Ical-\frac{\l(1{+}\m\mb)}{\l-\m}\,P \quad,
\end{equation}
whose location~$\m$ is a moduli parameter.
$P$ is a $\l$-independent string field to be determined.
Let us investigate for our ansatz~(\ref{ansatz1}) the consequences
of (\ref{nopole2}) and (\ref{nopole1}), in that order.
The residues of the $\l$-poles of $\Psi\ov{\Psi}$
at $\l{=}\m$ and $\l{=}-1/\mb$ are proportional to $\ P(\Ical{-}\ov{P})\ $
and $\ (\Ical{-}P)\ov{P}\ $ (for $\m\in\C P^1$ arbitrary and fixed),
respectively, implying the projector property
\begin{equation} \label{proj}
  P^2\=P\=\ov{P} \quad.
\end{equation}
This is achieved by parametrizing
\begin{equation} \label{PT}
  P \= T\,( \Tb T )^{-1}\Tb
\end{equation}
with some string field $T$.
Similarly, the absence of poles in~(\ref{nopole1}) yields
\begin{equation} \label{dP}
  P \bigl( \m Q + \h_0 \bigr) P \=0
  \qquad\textrm{and}\qquad
  (\Ical-P) \bigl( Q - \mb\h_0 \bigr) P \=0
\end{equation}
which are conjugate to one another.
Since $PT=T$ by construction these equations are satisfied if
\begin{equation} \label{dT}
  \bigl( Q - \mb\h_0 \bigr) T \= 0 \quad.
\end{equation}

It is important to note that $T$ is only subject
to a {\it linear\/} equation and otherwise unconstrained.
An obvious solution to~(\ref{dT}) is
\begin{equation} \label{solT}
  T \= \bigl( Q - \mb\h_0 \bigr) W
\end{equation}
for an arbitrary string field~$W$.
Every choice of~$W$ or solution to~(\ref{dT}) yields 
a classical Berkovits string field,
\begin{equation} \label{solPhi}
  \e^{-\P}\=\Ical-(1{+}\m\mb)\,P \quad,\qquad
  \e^{\P} \=\Ical-(1{+}\sfrac{1}{\m\mb})\,P
\end{equation}
and, from $\l\to0$,\footnote{
An alternative representation is
$A=-\sfrac{1{+}\m\mb}{\m\mb}\,\bigl[QP-PZ(\Ical{-}P)\bigr]$
where $Z$ is defined by $\bigl(Q-\mb\h_0\bigr)\Tb =: \Tb\,Z$.}
\begin{equation} \label{solA}
  A\=-\sfrac{1{+}\m\mb}{\m}\,\h_0\,P \quad.
\end{equation}

\noindent
{\bf 4.~Shifting the background.\ }
The form of the string field equations does not depend on the 
choice of background (termed ``vacuum''). However, the explicit
structure of the kinetic operator~$Q$ is determined by this choice.
For the open-string vacuum,
\begin{equation}
  A_0=0 \quad,\qquad P_0=0 \quad,\qquad \Psi_0=\Ical \quad,
\end{equation}
one has the familiar BRST operator, $Q=Q_{\text{B}}$.
Now, one may think of the solution $(\Psi,A)$ to~(\ref{linl})
as the result of a ``dressing map''~\cite{Lechtenfeld:2002cu}~\footnote{
We abbreviate ${\rm Ad}_\Psi A_0:=\Psi(Q+\sfrac{1}{\l}\h_0+A_0)\Psi^{-1}$.}
\begin{equation}
  \Psi_0=\Ical \quad\longmapsto\quad \Psi=\Psi(\l)\,\Psi_0
  \qquad\textrm{and}\qquad A_0=0 \quad\longmapsto\quad A={\rm Ad}_\Psi A_0
\end{equation}
applied to a ``seed solution'' $(\Psi_0,A_0)$.
This process can be iterated.
Since any two classical superstring configurations are related by
such a dressing transformation, a shift of the background
$(\Psi_0,A_0)$ to a new reference configuration~$(\Psi_1,A_1)$
is exactly of the same nature. The difference is only semantical.

We study the result of shifting the background 
by a dressing transformation according to
\begin{equation}
  \parbox{4cm}{
    \textrm{background:}
    \phantom{XXXXXXXXXXXXXXXXXX}
    \phantom{XXXXXXXXXXXXXXXXXX}
    \textrm{deviation:}
  }
  \begin{CD}
    \Psi_0{=}\Ical @>{\Psi_1}>> \Psi_1      \\
    @VV{\Psi}V                     @VV{\Psi'}V \\
    \Psi              @>>>         \Pst
  \end{CD}
  \qquad\qquad\qquad
  \begin{CD}
     A_0{=}0 @>{{\rm Ad}_{\Psi_1}}>> A_1                    \\
     @VV{{\rm Ad}_\Psi}V             @VV{{\rm Ad}_{\Psi'}}V \\
     A_0{+}A @>>>                    \At
  \end{CD}
\end{equation}
where horizontal arrows represent the dressing map to the new background
and vertical arrows turn on a deviation via dressing. Composing the two
dressing transformations, the linear equation becomes
($\Pst=\Psi'\Psi_1$ and $\At=A_1{+}A'$)
\begin{eqnarray}
  \nonumber 0 &\=&
    \bigl( Q + \sfrac{1}{\l}\h_0 + \At \bigr)\,\Pst \\[8pt]
  \nonumber   &\=&
    \bigl[ Q \Psi' + A_1 \Psi' - \Psi' A_1 + \sfrac{1}{\l}\h_0
    \Psi' + (\At-A_1)\Psi' \bigr] \,\Psi_1 \\[8pt]
  &\=& \bigl[ \bigl( Q' + \sfrac{1}{\l}\h_0 + A' \bigr)\,
    \Psi'\bigr]\,\Psi_1 \quad,
\end{eqnarray}
where we used $\ (Q+\frac{1}{\l}\h_0)\Psi_1=-A_1\Psi_1\ $ and defined
$\ Q'\Psi':=Q\,\Psi'+A_1\Psi'-\Psi'A_1$.
Hence, measuring our string fields from the new vacuum~$A_1$,
the relevant linear system,
\begin{equation} \label{lint}
  \bigl( Q' + \sfrac{1}{\l}\h_0 + A' \bigr)\,\Psi' \=0 \quad,
\end{equation}
has the same form as the original~(\ref{linl}), but $Q$ has changed
into~$Q'$. For the nonlinear string field equations the corresponding
form invariance has been observed in~\cite{Marino:2001ny}, a fact
almost trivial in our framework.

\noindent
{\bf 5.~Tachyon vacuum superstring fields.\ }
Of special interest is the form of the theory around the tachyonic vacuum.
Deviations from the tachyon vacuum are governed by (\ref{lint}),
and all equations pertaining to the open-string vacuum simply carry over
(with primes added).
However, this is not the whole story. As discussed in~%
\cite{Gaiotto:2001ji, Ohmori:2002kj}, a new kinetic operator
built entirely from ghosts can be ``derived'' via a redefinition 
of the new (tachyon vacuum) superstring fields, 
\begin{equation} \label{redef}
  A'\ \mapsto\ \Ucal_{\ve_r}\,A'\ =:\ \Ach
  \qquad\textrm{and}\qquad
  \Psi'\ \mapsto\ \Ucal_{\ve_r}\,\Psi'\ =:\ \Psh \quad,
\end{equation}
such that
\begin{equation}
  Q'\ \mapsto\ \Ucal_{\ve_r}\,Q'\,\Ucal_{\ve_r}^{-1}\ =:\ \Qch
\end{equation}
yields the proper zero-cohomology ``vacuum'' kinetic operator.
The field redefinition~(\ref{redef}) is induced by a world-sheet
reparametrization which is singular for $\ve_r\to0$.
As $\h$ has conformal spin one, its zero mode~$\h_0$ is inert under the
reparametrization. From now on, a hat indicates the presence of internal 
$2{\times}2$ Chan-Paton matrices distinguishing the GSO$(\pm)$ sectors, e.~g.,
\begin{align}
  \Ach & \= \Acal_+\otimes \s_3 + \Acal_-\otimes \i\s_2
    & & \mbox{(odd ghost number)} \quad, \\
  \Phh & \= \P_+\otimes \boldsymbol{1} + \P_-\otimes \s_1
    & & \mbox{(even ghost number)} \quad.
\end{align}

The kinetic operator of this ``vacuum superstring field theory'' (VSSFT)
is conjectured to have the form~\cite{Arefeva:2002mb, Ohmori:2002kj}
\begin{equation} \label{Qone}
  \Qch\=\Qcal_{\text{odd}}\otimes\s_3+\Qcal_{\text{even}}\otimes\i\s_2 \quad,
\end{equation}
where the subscript refers to the Grassmann parity and
\begin{eqnarray} \label{Qtwo}
  \Qcal_{\text{odd}}&\=&\sfrac{1}{4\i\ve_r^2}\bigl[c(\i)-c(-\i)\bigr]
  \ +\ {\textstyle \oint} \tfrac{dz}{2\pi\i} b\g^2(z) \quad, 
  \\[8pt] \label{Qthree}
  \Qcal_{\text{even}}&\=&\sfrac{1}{2\i\ve_r}\bigl[\g(\i)-\g(-\i)\bigr]
  \,\Pi_+ \ +\ \sfrac{1}{2\ve_r}\bigl[\g(\i)+\g(-\i)\bigr]\,\Pi_-
\end{eqnarray}
with projectors $\Pi_+$ and $\Pi_-$ onto the GSO$(+)$ and GSO$(-)$ sectors, 
respectively. These terms prevail in the limit $\ve_r\to0$.
Consequently, the linear system for VSSFT reads
\begin{equation} \label{linv}
  \bigl( \Qch + \sfrac{1}{\l}\hh + \Ach \bigr)\,\Psh(\l) \=0 \quad,
\end{equation}
where $\ \hh=\h_0\otimes \s_3\ $ and 
$\ \Psh=\Psi_+\otimes\boldsymbol{1} + \Psi_-\otimes \s_1$.
Again, solutions to Berkovits' VSSFT or to the cubic VSSFT are obtained
from (\ref{solPhi}) or~(\ref{solA}) by firstly solving the linear equation~%
(\ref{dT}) after replacing $Q\to\Qch$ and secondly composing the projector
via~(\ref{PT}).

It is usually assumed that the D-brane solutions of VSSFT factorize
into a ghost and a matter part, $\Ach=\Ahg\otimes\Acal_m$. 
Then, the cubic VSSFT equation,
\begin{equation} \label{ceom}
  \Qch\,\Ach + {\Ach}^2\=0 \qquad\textrm{with}\qquad
  \hh\,\Ach\=0 \quad,
\end{equation}
splits into
\begin{equation} \label{F0g}
  \Acal_m^2\=\Acal_m \qquad\textrm{and}\qquad
  \Qch\,\Ahg + {\Ahg}^2 \=0 \quad\textrm{with}\quad \hh\,\Ahg\=0
\end{equation}
which turns $\Acal_m$ into a projector.
Within our single-pole ansatz~(\ref{ansatz1}), the full $\Ach$ 
is already proportional to a projector~$\ \Pch=\Phg\otimes\Pcal_m$, 
hence we must simply factorize~(\ref{solA}) and have
\begin{equation} \label{AP}
  \Acal_m = \Pcal_m \quad\textrm{and}\quad
  \Ahg = -\sfrac{1{+}\m\mb}{\m}\,\hh\,\Phg \qquad\textrm{with}\qquad
  \Pcal_m^2=\Pcal_m \quad\textrm{and}\quad
  \Phg^2 = \Phg \quad.
\end{equation}
Since $\Qch$ is pure ghost the projector equation~(\ref{dP}) factorizes,
and (\ref{F0g}) reduces to (\ref{AP}) plus
\begin{equation} \label{projg}
  (\wh{\Ical}_g - \Phg)\bigl( \Qch - \mb \hh \bigr) \Phg \= 0 \quad,
\end{equation}
which is solved by (we omit hats over~$\Tcal_g$)
\begin{equation} \label{Tg}
  \Phg \= \Tcal_g\,(\Tcb_g \Tcal_g)^{-1}\Tcb_g \qquad\textrm{and}\qquad
  \bigl( \Qch - \mb \hh \bigr) \Tcal_g \= 0 \quad .
\end{equation}
  
In the nonpolynomial formulation, a different ansatz, 
$\Phh=\Phh_g\otimes\P_m\ $ with $\ \P_m^2=\P_m$, 
was advocated by Mari\~{n}o and Schiappa~\cite{Marino:2001ny}. 
It allows one to factorize Berkovits' equation~(\ref{yang}) since one gets
\begin{equation}
  \e^{\pm\Phh}\=
  \wh{\Ical}-\bigl(\wh{\Ical}_g-\e^{\pm\Phh_g}\bigr)\otimes\P_m \=
  \wh{\Ical}_g\otimes\bigl(\Ical_m{-}\P_m\bigr) +
  \e^{\pm\Phh_g} \otimes \P_m \quad.
\end{equation}
However, comparison with our solution~(\ref{solPhi}),
\begin{equation}
  \e^{\pm\Phh}\=
  \wh{\Ical}-\bigl( 1{+}(\m\mb)^{\mp1} \bigr)\,\Phg \otimes \Pcal_m \quad,
\end{equation}
implies $\ \P_m=\Pcal_m\ $ and $\ \Phh_g=-(\ln\m\mb+\i\pi)\,\Phg\ $
which is not compatible with the reality of~$\P$. Hence, our ansatz
differs from the one of~\cite{Marino:2001ny}.

A more important distinction of our single-pole ansatz~(\ref{ansatz1})
from previous work is visible from~(\ref{Tg}):
The cohomology problem for $\Tcal_g$ is not based on $\Qch$ but on 
$\Qch{-}\mb\hh$. Motivated by the freedom to choose a particular
embedding of an \N2 superconformal algebra into a small \Ng4
superconformal algebra, such a coboundary operator (in the case of
the open string vacuum) was proposed initially in~%
\cite{Berkovits:1994vy,Berkovits:1999im}.

\noindent
{\bf 6.~Ghost picture modification.\ }
As it stands, the linear equations~(\ref{linv}) and~(\ref{Tg}) face a
problem due to the ghost picture degeneracy of the NSR~superstring. If
our string fields are to carry a definite picture charge, they must
reside in the zero-picture sector. Since $\h_0$ lowers the picture charge
by one unit, the above-mentioned coboundary operator is not homogeneous
in picture. Therefore, from~(\ref{linv}) or~(\ref{Tg}) one concludes
that any string field, including $\Ach$ and~$\Tcal_g$, must in general be
an infinite sum over all picture sectors. 
Obviously, any such field may be expanded into a formal series
$\ \Tcal_g=\sum_{n\in\Z}(-\mb)^{-n}\Tcal_n$,
where $\Tcal_n$ carries picture number $n$. From~(\ref{Tg})
we then obtain the recursion relations $\ \hh\Tcal_{n+1}=-\Qch\Tcal_n$.
If we want to maintain Berkovits' original proposal that all string
fields have picture number zero (e.~g., $\Tcal_{n\neq 0}=0$) then only the
trivial solutions of~(\ref{linv}) with $\ \Qch \Tcal_0 = 0 = \hh \Tcal_0\ $ 
emerge. Clearly, this implies $\ \Qch\,\Phg = 0 = \hh\,\Phg\ $
and therefore $\ \Ach=0$. The supersliver~\cite{Marino:2001ny,Arefeva:2001ke} 
is gauge equivalent to this vacuum~\cite{Ohmori:2002ah}. 

To obtain nontrivial solutions, we have two possibilites: Either we
admit string fields inhomogeneous in picture, or we modify our linear
equation. In the following we shall pursue the second option and 
restrict all string fields to the zero picture.
The obvious cure then is to introduce a picture-raising multiplier,
$\hh\to \Xch(\i)\hh$.
This is admissible as long as $\Xch(\i)$ commutes with both $\hh$
and~$\Qch$ and can be pulled through the star product.\footnote{
Any midpoint insertion of conformal spin zero commutes with Witten's
star product, as can be seen by its definition in terms of correlation
functions of the disk.}
We propose to take $\ \Xch(\i):=\{\Qch,\wh{\xi}(\i)\}$,
i.~e. the picture-raising operator~$\Xch$ of VSSFT evaluated at the
string midpoint.\footnote{Due to the explicit form~(\ref{Qone})--(\ref{Qthree})
of the kinetic operator, $\Xch(\i)$ consists of Grassmann-even and -odd
parts. The Grassmann-even part simply reads $-\pa(b\h\e^{2\phi})(\i)
-b\pa\h\e^{2\phi}(\i)$; the Grassmann-odd part has to be regularized
due to the pole in the OPE of $\g$ with $\xi$. Around the
open-string vacuum, we may simply take $X(\i)=\{Q,\xi(\i)\}$.}
With this modification, our master linear equation becomes
\begin{equation} \label{linX}
  \bigl( \Qch + \sfrac{1}{\l}\Xch(\i)\hh + \Ach \bigr)\,\Psh(\l) \=0 \quad,
\end{equation}
and all subsequent equations continue to hold after the obvious
insertions of $\Xch(\i)$. In particular, the ghost picture modification 
changes Berkovits' string field equation~(\ref{yang}) to
\begin{equation}
  \Xch(\i)\,\hh \bigl( \e^{-\Phh}\Qch\,\e^{\Phh} \bigr) \=0 \quad .
\end{equation}
Any solution~$\Ach$ in the form of~(\ref{solA}) will, however,
automatically be annihilated by $\hh$ so that it fulfills also
Berkovits' equation of motion without~$\Xch(\i)$. Note that the action
will remain unchanged; we use $\Xch(\i)$ only as a means to solve our
linear equations.

\noindent
{\bf 7.~Towards explicit solutions.\ }
In order to extract the physical properties of classical VSSFT configurations,
e.~g., a D-brane interpretation or the role of our moduli parameter~$\m$, 
it is desirable to construct solutions to the field equations 
in a more explicit manner.
In keeping with the paradigm of matter-ghost factorization (see (\ref{F0g}))
we are asked to solve eq.~(\ref{Tg}) with $\Xch(\i)$ inserted.
Because $\ \Qch-\mb\Xch(\i)\hh\ $ can be ``inverted''
the general solution of VSSFT may be constructed from
\begin{equation} \label{solTc}
  \Tcal_g \= \bigl( \Qch - \mb\Xch(\i)\hh \bigr)\,\wh{\Wcal}_g
\end{equation}
for an arbitrary ghost string field~$\wh{\Wcal}_g$.

For cubic VSSFT, the $\ve_r$ expansion of~\cite{Ohmori:2002kj} can be
reproduced in this framework.\footnote{
Our $\Qcal_{\text{odd}}$ in~(\ref{Qtwo}) has a different $\ve_r$-dependence
(coinciding with that of~\cite{Arefeva:2002mb}), but this is irrelevant here.}
In particular, since the leading term of $\ \Qch{-}\mb\Xch(\i)\hh\ $
is identical to $\ \Qcal_{\text{GRSZ}}{\otimes}\s_3\ $~\cite{Gaiotto:2001ji},
the lowest order in $\ve_r$ involves only the ``natural'' Grassmann assignments
of all quantities.

Certain special solutions can be seen directly.
When $\mb=1$, for instance, one may employ the picture-lowering operator~%
$\wh{Y}(\i)$ to write $\ \Tcal_g = \wh{Y}(\i)\,\wh{\xi}(\i)\,\wh{\Xi}_g\ $
where $\ \Qch\,\wh{\Xi}_g=0=\hh\,\wh{\Xi}_g$. 
At leading order in~$\ve_r$ we may identify $\wh{\Xi}_g$ with 
the ghost supersliver~$\Xi_g{\otimes}{\bf1}$.

In any case, the main difficulty arises in the composition of $\Pch_g$
from a given~$\Tcal_g$ since Witten's star product is implicit in~(\ref{Tg}).
In order to circumvent this technical obstacle we propose to make use of the 
(discrete~\cite{Bars:2001ag} or continuous~\cite{Douglas:2002jm}) 
Moyal formulation of Witten's star product.
In such a situation, the Moyal-Weyl map can be inferred to encode the
non(anti)commutativity into Heisenberg or Clifford algebras, which 
are represented in auxiliary Fock spaces.\footnote{ 
Such Fock spaces are not to be confused with the string oscillator Fock space.}
The advantage of this (auxiliary) operator formulation is 
its calculational ease. As an example, 
the basic projector for a single Moyal pair can be expressed as follows:
\begin{eqnarray} \label{P0b}
[a,a^\+]=1 \quad & \Longrightarrow & \quad
|0\>\<0| \=\ :\e^{-a^\+ a}:\ 
%        \= \bigl[ 2\,\e^{-2a^\+ a} \bigr]\big|_\text{Weyl-ordered}
         \= 1 - a^\+ (a a^\+)^{-1} a \quad, \\[10pt] \label{P0f}
\{c,c^\+\}=1 \quad & \Longrightarrow & \quad
|0\>\<0| \=\ :\e^{-c^\+ c}:\
	 \= 1 - c^\+c
         \= 1 - c^\+ (c\,c^\+)^{-1} c \quad,
\end{eqnarray}
displaying a simple connection between the Gaussian form and 
the ``fractional'' form (cf.~(\ref{PT})) of a projector. 
Of course, for the application to VSSFT infinite tensor products of 
Heisenberg and Clifford algebras have to be considered~%
\cite{Bars:2001ag,Douglas:2002jm,Arefeva:2002jj,Erler:2002nr}.
However, (\ref{P0b}) and (\ref{P0f}) suggest the possibility to take $\Tcal_g$
not to be an operator but a state $|\Tcal_g\>$ in the auxiliary Fock space.
This would be in tune with the construction of 
noncommutative abelian solitons~\cite{Lechtenfeld:2001aw}.
Finally, a direct comparison with results in the conventional string 
oscillator basis requires the reverse basis transformation to be applied
to the string field configurations constructed in the Moyal basis.
%\goodbreak

In closing, we should like to stress that we have reduced the problem of
solving the superstring field equations (in cubic or nonpolynomial form)
to the easier task of considering a linear equation, whose solution~$T$
then serves as a building block for the string field configuration.
Although demonstrated here with the simplest (single-pole) ansatz for
the auxiliary string field~$\Psi(\l)$, this strategy generalizes to the 
universal (multi-pole) case~\cite{Lechtenfeld:2002cu}. Projectors emerge
naturally only in the single-pole setup while $T$ (rather a collection of such)
continues to play the decisive role.  The formalism is ideally suited 
to handle the superposition of solitonic objects in integrable systems.
We therefore expect it to yield multi-brane configurations automatically.

\bigskip
\noindent
{\large{\bf Acknowledgements}}

\smallskip\noindent
O.~L.\ acknowledges discussions with I.~Aref'eva and L.~Bonora.
This work is partially supported by DFG grant Le~838/7-1 in the priority
program ``String Theory'' (SPP 1096).

\end{document}